\documentclass{nature}

\usepackage{graphicx}
\usepackage{amssymb,amsmath}
\usepackage{times}
\usepackage{mathptmx}
\def\apj{{Astrophys. J.}}
\def\aap{{Astron. \& Astrophys.}}
\def\aj{{Astron. J.}}
\def\pasp{{Pub. Astron. Soc. Pac.}}
\def\apjl{{Astrophys. J. Let.}}

\def\mnras{{Mon. Not. R. Astron. Soc.}}
\def\araa{{Ann. Rev. Astron. \& Astrophys.}}

\def\ssr{{\rm Space~Sci.~Rev.}} 

\title{Supernova 2011fe from an exploding carbon-oxygen white dwarf star}

\author{Peter E. Nugent$^{1,2}$, Mark Sullivan$^3$, S. Bradley
  Cenko$^2$, Rollin C. Thomas$^1$, Daniel Kasen$^{1,4}$, D. Andrew
  Howell$^{5,6}$, David Bersier$^7$, Joshua S. Bloom$^2$, S.~R.
  Kulkarni$^8$, Michael T. Kandrashoff$^2$, Alexei V.  Filippenko$^2$,
  Jeffrey M. Silverman$^2$, Geoffrey W. Marcy$^2$, Andrew W.
  Howard$^2$, Howard T.  Isaacson$^2$, Kate Maguire$^3$, Nao
  Suzuki$^1$, James E.  Tarlton$^3$, Yen-Chen Pan$^3$, Lars
  Bildsten$^{6,9}$, Benjamin J.  Fulton$^{5,6}$, Jerod T.
  Parrent$^{5,10}$, David Sand$^{5,6}$, Philipp Podsiadlowski$^3$,
  Federica B.  Bianco$^{5,6}$, Benjamin Dilday$^{5,6}$, Melissa L.
  Graham$^{5,6}$, Joe Lyman$^7$, Phil James$^7$, Mansi M.
  Kasliwal$^8$, Nicholas M. Law$^{11}$, Robert M.  Quimby$^12$, Isobel
  M. Hook$^{3,13}$, Emma S. Walker$^{14}$, Paolo Mazzali$^{15,16}$,
  Elena Pian$^{15,16}$, Eran O.  Ofek$^{8,17}$, Avishay Gal-Yam$^{17}$
  and Dovi Poznanski$^{1,2,18}$}

\begin{document}

\maketitle

\begin{affiliations}
\item Lawrence Berkeley National Laboratory, Berkeley, CA, 94720, USA.
\item Department of Astronomy, University of California, Berkeley, CA,
  94720-3411, USA. 
\item Department of Physics (Astrophysics), University of Oxford,
  Keble Road, Oxford, OX1 3RH, UK. 
\item Department of Physics, University of California, Berkeley, CA,
  94720, USA. 
\item Las Cumbres Observatory Global Telescope Network, Goleta, CA,
  93117, USA. 
\item Department of Physics, University of California, Santa Barbara,
  CA, 93106, USA.
\item Astrophysics Research Institute, Liverpool John Moores
  University, Birkenhead, UK.
\item Cahill Center for Astrophysics, California Institute of
Technology, Pasadena, CA, 91125, USA.
\item Kavli Institute for Theoretical Physics, University of
  California, Santa Barbara, CA, 93106, USA.
\item Department of Physics and Astronomy, Dartmouth College, Hanover,
  NH, USA.
\item Dunlap Institute for Astronomy and Astrophysics, University of
Toronto, 50 St. George Street, Toronto M5S 3H4, Ontario, Canada.
\item IPMU, University of Tokyo, Kashiwanoha 5-1-5, Kashiwa-shi,
  Chiba, Japan
\item INAF, Osservatorio Astronomico di Roma, via Frascati 33, 00040
  Monteporzio (RM), Italy 
\item Scuola Normale Superiore, Piazza dei Cavalieri 7, 56126 Pisa,
  Italy
\item INAF-Osservatorio Astronomico di Padova, vicolo
  dell'Osservatorio, 5 35122 Padova, Italy
\item Max-Planck Institut fuer Astrophysik Karl-Schwarzschildstr. 1
  85748 Garching, Germany
\item Benoziyo Center for Astrophysics, The Weizmann Institute of
  Science, Rehovot 76100, Israel
\item School of Physics and Astronomy, Tel-Aviv University, Tel-Aviv
  69978, Israel 
\end{affiliations}

\clearpage

\begin{abstract}
  Type Ia supernovae (SNe~Ia) have been used empirically as
  standardized candles to reveal the accelerating
  universe\cite{1998AJ....116.1009R, 1999ApJ...517..565P,
    2011ApJ...737..102S} even though fundamental details, such as the
  nature of the progenitor system and how the star explodes, remained
  a
  mystery\cite{2011NatCo...2E.350H,2009Natur.460..869K,2007MNRAS.380..933Y}.
  There is consensus that a white dwarf star explodes after accreting
  matter in a binary system, but the secondary could be anything from
  a main sequence star to a red giant, or even another white dwarf.
  The uncertainty stems from the fact that no recent SN~Ia has been
  discovered close enough to detect the stars before explosion.  Here
  we report early observations of SN~2011fe (PTF11kly) in M101 at a
  distance of 6.4~Mpc\cite{2011ApJ...733..124S}, the closest SN~Ia in
  the past 25 years. We find that the exploding star was likely a
  carbon-oxygen white dwarf, and from the lack of an early shock we
  conclude that the companion was most likely a main sequence star.
  Early spectroscopy shows high-velocity oxygen that varies on a time
  scale of hours and extensive mixing of newly synthesized
  intermediate mass elements in the outermost layers of the supernova.
  A companion paper\cite{li_nature} uses pre-explosion images to rule
  out luminous red giants and most helium stars as companions.
\end{abstract}

\begin{figure}[t]
\begin{center}
\includegraphics[width=6.5in]{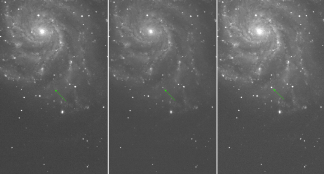}
  \caption{\label{fig:disc} \normalsize{PTF $g$-band image sequence of
      the field of Messier 101 showing the appearance of SN~2011fe.
      From left to right, images are from August 23.22, 24.17, and
      25.16 UT. The supernova was not detected on the first night to a
      3-$\sigma$ limiting magnitude of 21.5, was discovered at
      magnitude 17.35, and increased by a factor of 10 in brightness
      to mag 14.86 the following night. The supernova peaked
      at magnitude $\sim$9.9, making it the fifth
      brightest supernova in the past century. PTF is a wide-field
      optical experiment designed to systematically explore the
      variable sky on a variety of time scales, with one particular
      focus the very early detection of SNe\cite{2009PASP..121.1334R,
        2009PASP..121.1395L}.  Discoveries such as this one have been
      made possible by coupling real-time computational tools to
      extensive astronomical follow-up
      observations\cite{2011ApJ...736..159G,brn+11}.}}
\end{center}
\end{figure}

SN~2011fe was detected on 2011 August 24.167 (UT 03:59) with a
$g$-band magnitude of 17.35 by the Palomar Transient Factory (PTF) in
the Pinwheel galaxy --- Messier 101 (M101; see Figure~\ref{fig:disc}).
Observations on the previous night revealed no source to a limiting
magnitude of 21.5.  Given the distance to M101 of
6.4~Mpc\cite{2011ApJ...733..124S}, this first observation identified
the supernova at an absolute magnitude of $-11.7$, roughly 1/1000 of
its peak brightness.

Following an alert sent to the PTF consortium (at UT 19:51),
observations were immediately undertaken by the {\it Swift}
Observatory, and spectroscopic observations were carried out at UT
20:42 on the robotic Liverpool Telescope equipped with the FRODOSpec
spectrograph (located at La~Palma, Canary Islands).  After the
calibration of this spectrum, at UT 23:47 an Astronomer's Telegram was
issued\cite{2011ATel.3581....1N} identifying SN~2011fe as a young
supernova of Type Ia.  Eight hours later, a low-resolution spectrum
was obtained with the Kast spectrograph at the Lick 3-m Shane
telescope (Mt.\ Hamilton, California) and a high-resolution spectrum
with HIRES at the Keck~I telescope (Mauna Kea, Hawaii). These spectra
are presented in Figure~\ref{fig:spec} (see Supplementary Information
for details).

The discovery and extensive follow-up photometry allow us to estimate
the time of explosion to high precision (Figure~\ref{fig:rise}).  At
very early times the luminosity should scale as the surface area of
the expanding fireball, thus is expected to rise as $t^2$, where $t$
is the time since explosion. This assumes neither the photospheric
temperature nor the velocity change significantly and the input energy
from the radioactive decay of $^{56}$Ni to $^{56}$Co is relatively
constant over this period and is near the photosphere.  {\it Swift}
observations show only small changes in the relative flux between the
optical and ultraviolet, and the velocity evolution over the first
24\,hr is small --- consistent with these assumptions (see
Supplementary Information).

Using the $t^2$ model, we find an explosion time at modified Julian
date (MJD)\,$55796.696\pm0.003$ (see Figure~\ref{fig:rise}). Letting
the exponent of the power law depart from 2, which captures some of
the deviations from the fireball model, and fitting just the first 3
nights of data, results in a best-fit explosion date of
$55796.687\pm0.014$ (UT 2011 August 23, 16:29 $\pm$ 20 minutes). The
exponent of the power law is $2.01\pm0.01$, consistent with the model
discussed above.  Based on these fits, our first data points were
obtained just over 11\,hr after SN~2011fe exploded.

We analysed the Lick spectrum of SN~2011fe using the automated SN
spectrum interpretation code SYNAPPS\cite{2011PASP..123..237T} (see
Supplementary Information for further details). At this time only a
few hundredths of a solar mass of material are visible above the
photosphere, yet typical\cite{1997ARA&A..35..309F} pre-maximum SN~Ia
ions are seen: O~I, Mg~II, Si~II, S~II, Ca~II, and Fe~II are present
at velocities of 16,000 km\,s$^{-1}$. The fit also shows the presence
of C~II $\lambda\lambda$6580, 7234.  Fe~III was not needed in the fit.
Both high-velocity (HV) Si~II and Ca~II are confirmed by SYNAPPS
(extending above 21,000 km\,s$^{-1}$).  Surprisingly, SYNAPPS finds HV
O~I (in excess of 20,000 km\,s$^{-1}$) for the absorption centered at
7400~\AA. This feature has evolved significantly in only 8\,hr,
between the data taken at the Liverpool Telescope and Lick, with the
absorption minimum receding from 18,000 to 14,000~km\,s$^{-1}$. The
rapid evolution in these optically thin layers is best explained by
geometrical dilution during the early phases. To our knowledge, this
is the first identification of rapidly evolving high-velocity oxygen
in the ejecta of a SN~Ia.

The early-time spectra provide fundamental insight into the explosion
physics of this SN.  As in previous\cite{1983ApJ...270..123B} SNe~Ia,
intermediate-mass elements dominate the spectrum. In addition, we see
strong features from unburnt material (carbon and high-velocity
oxygen).  The overlap in velocity space implies that the explosion
processed the outer layers of the progenitor white dwarf, but left
behind (at least some) carbon and oxygen.  The unburnt material could
be confined to pockets, or the ejecta in the outer layers may be
thoroughly mixed.  The doubly ionised species (e.g., Si~III and
Fe~III), often seen in the spectra of many early and maximum-light
SNe~Ia\cite{1991ApJ...371L..23L,1995ApJ...455L.147N}, are absent even
though our observations are at $\sim 1$ day after explosion when the
energy input from radioactive decay is near its peak. SN~2011fe is
spectroscopically most similar to the slightly underluminous SNe~Ia
1992A and 1994D\cite{1993ApJ...415..589K,1996MNRAS.278..111P}, the
latter of which also shows high-velocity features in the $-12$ day
spectrum\cite{1999ApJ...525..881H}. One potential explanation for this
is that while some $^{56}$Ni has been mixed out to the photosphere,
the majority produced in the explosion is confined to the innermost
layers of the atmosphere, and thus the bulk of the heating is well
separated from the portion of the atmosphere we view in these spectra.

The early detection of SN~2011fe allows us to put considerable
constraints on the progenitor system of this SN~Ia. At early times
(about a day or less after explosion), radiative diffusion from the
shock-heated outer layers of the ejecta is a contributor to the
supernova luminosity.  The origin of the shock can either be from a
detonation of the WD\cite{piro_2010} or a later collision with the
companion star\cite{2010ApJ...708.1025K}.  Dimensionally, the shock
luminosity in this cooling, expanding envelope phase is $L \propto
E(t)/t_d$, where $E(t)$ is the ejecta internal energy at the elapsed
time $t$ and $t_d$ is the effective diffusion time through the
homologously expanding remnant.  Since the ejecta in these phases are
heavily radiation dominated, the internal energy declines during
adiabatic expansion as $E(t) \propto R_0/vt$, where $R_0$ is the
initial radius of the star.  Thus, the early-time luminosity is
proportional to $R_0$, while the effective temperature
$T_{\mathrm{eff}} \propto L^{1/4} \propto R_0^{1/4}$ (see also
Supplementary Information).

While there must be radioactive heating in the outer layers of the SN,
we can make the very conservative upper-limit approximation that the
earliest $g$-band photometric point ($L \approx 10^{40}\,{\rm
  erg~s^{-1}}$ at $\sim 0.5$~day) is {\it entirely} due to the
explosion. We then infer an upper limit to the radius of the
progenitor star, $R_0 < 0.1\,{\rm R}_\odot$
(Figure~\ref{fig:early_lc}).  This provides compelling, direct
evidence that the progenitor of SN~2011fe was a compact star, namely a
white dwarf.  When we add the early carbon and oxygen observations, we
conclude that the progenitor must have been a carbon-oxygen white
dwarf.

The early-time light curve also constrains the properties of a binary
star system\cite{2010ApJ...708.1025K}, as the collision of the SN with
a companion star will shock and reheat a portion of the ejecta.  The
resulting luminosity is proportional to the separation distance, $a$,
between the stars, and will be most prominent for observers aligned
with the symmetry axis.  A red-giant companion predicts an early
luminosity several orders of magnitude greater than that observed, and
can be ruled out regardless of the viewing angle.  A main-sequence
companion is compatible with the data, unless SN~2011fe happened to be
seen on-axis (within $\sim 40^\circ$ of the symmetry axis), in which
case the luminosity at day 0.5 rules out any binary with $a \leq
0.1\,{\rm R}_\odot$.

Recent simulations of double-degenerate mergers have found that some
material from the disrupted secondary WD may get pushed out to large
radius ($10^{13}$--$10^{14}$\,cm), either in the dynamics of the
merger\cite{2010ApJ...725..296F} or in the subsequent long-term
thermal evolution of the system\cite{2011arXiv1108.4036S}.  The
interaction of the ejecta with this roughly spherical medium should
produce (for all viewing angles) bright, early UV/optical emission, in
conflict with what is observed.  Our restriction that the dense
circumstellar medium must reside at $\leq 10^{10}$\,cm thus presents a
tight constraint for merger models, and only a few of those proposed
thus far may be allowable for this SN\cite{2007MNRAS.380..933Y}.

Using some of the earliest photometry and spectroscopy ever obtained
for a SN Ia, we have put more stringent limits than ever before
achieved for a SN Ia on the progenitor of SN~2011fe, revealing that
the primary is a carbon-oxygen white dwarf and the secondary is most
consistent with being a main sequence star.  We caution that these
constraints rely on theoretical interpretation.  A companion
letter\cite{li_nature} uses independent methodology to place direct
observational limits on the companion star from historical imaging
$\sim 100$ times deeper than previous attempts.

These results are from only the first $\sim $ week of observation of
SN~2011fe.  This first close SN Ia in the era of modern
instrumentation will undoubtedly become the best-studied thermonuclear
supernova in history allowing daily study from the UV to the IR well
into the faint nebular phase.  As such, it will form the new
foundation upon which our knowledge of more distant Type Ia supernovae
is built.

\clearpage

{\bf Acknowledgments}: 

The Palomar Transient Factory project is a scientific collaboration
between the California Institute of Technology, Columbia University,
Las Cumbres Observatory, the Lawrence Berkeley National Laboratory,
the National Energy Research Scientific Computing Center, the
University of Oxford, and the Weizmann Institute of Science.  The
National Energy Research Scientific Computing Center, supported by the
Office of Science of the U.S. Department of Energy, provided staff,
computational resources, and data storage for this project.  P.N.
acknowledges support from the US DOE Scientific Discovery through
Advanced Computing program. M.S. acknowledges support from the Royal
Society.  J.S.B. \& L.B. were supported by NSF.  The work of A.V.F. is
funded by NSF, the TABASGO Foundation, Gary and Cynthia Bengier, and
the Richard and Rhoda Goldman Fund. A.G. thanks the ISF and BSF. The
Liverpool Telescope is operated by Liverpool John Moores University in
the Spanish Observatorio del Roque de los Muchachos of the Instituto
de Astrofisica de Canarias with financial support from the UK Science
and Technology Facilities Council. Some of the data presented herein
were obtained at the W. M.  Keck Observatory, which is operated as a
scientific partnership among the California Institute of Technology,
the University of California, and NASA; the observatory was made
possible by the generous financial support of the W. M. Keck
Foundation. We thank the staffs of the many observatories at which
data were obtained for their excellent assistance.

{\bf Author Contributions}:

PEN, MS, and DAH oversee the PTF SN Ia program.  PEN oversaw the
preparation of the manuscript.  MS, DB, KM,
YCP, JL, and PJ executed and reduced the FRODOSpec observations.  SBC,
MTK, AVF, and JMS executed and reduced the Lick spectrum.  GWM, AWH,
and HTI obtained the HIRES observations.  SBC, JSB, SRK, MMK, NML,
EOO, RMQ, and DP assisted in the operation of P48 as part of the PTF
collaboration.  RCT and JET performed the SYNAPPS analysis.  DK, LB,
and PP assisted with the theoretical interpretation of our
observations.  NS, BJF, JTP, DS, FB, BD, MLG, IMH, PM, EP, EW, AG
assisted in follow-up observations of SN~2011fe.

Correspondence and requests for materials should be addressed to Peter
Nugent (e-mail: penugent@lbl.gov).

\clearpage
\thispagestyle{empty}

\clearpage
\thispagestyle{empty}

\begin{figure}[t]
\begin{center}
 \includegraphics[width=6.5in]{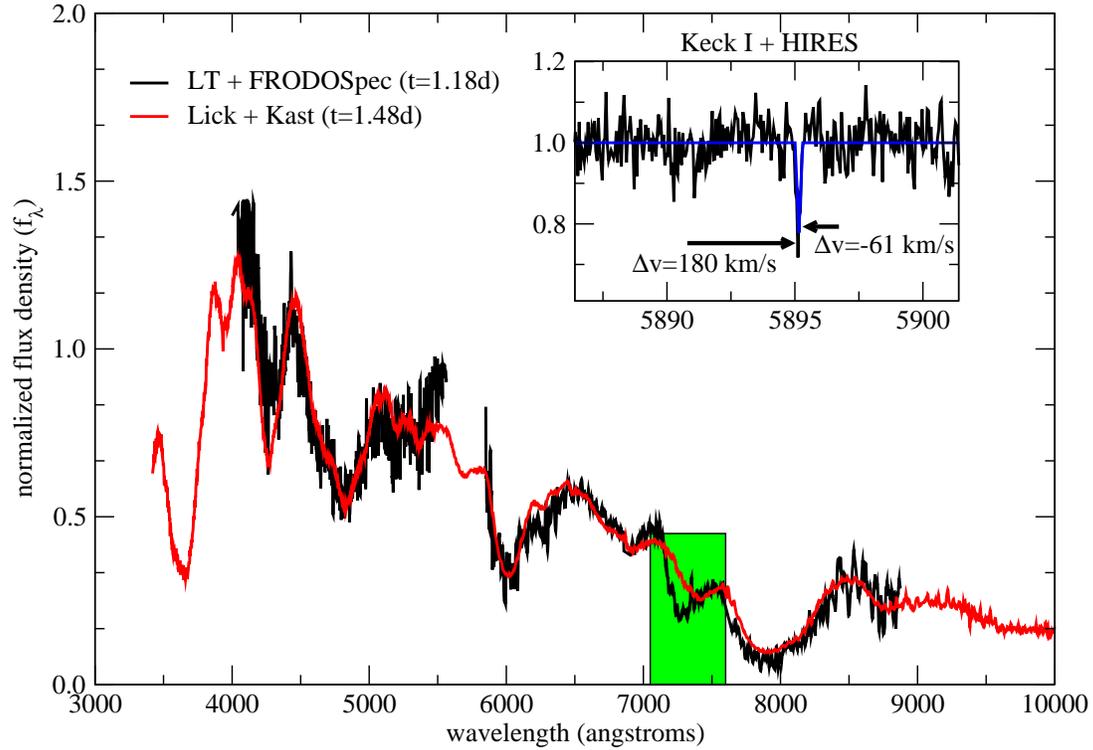}
  \caption{\label{fig:spec} \footnotesize{Spectra of SN~2011fe taken
      1.5 days after explosion. Typical pre-maximum
      SN~Ia ions are seen: O~I, Mg~II, Si~II, S~II, Ca~II, and Fe~II
      are present at photospheric velocities of 16,000 km\,s$^{-1}$.
      In addition, the fit shows the presence of C~II
      $\lambda\lambda$6580, 7234.  Both high-velocity (HV) Si~II and
      Ca~II are seen (extending above 21,000 km\,s$^{-1}$), as is HV
      O~I, the first evidence of such a feature in a SN~Ia (green
      highlighted region). Note that this feature evolves in $\sim
      8$\,hr between the first two low-resolution spectra. {\it
        Inset:} A Keck+HIRES spectrum centered on the Na~I~D line. In
      this wavelength range, we identify only a single significant
      absorption feature.  Fitting a Gaussian profile to it, we
      measure a central wavelength of $\lambda = 5893.75 \pm
      0.02$\,\AA\ and a full width at half-maximum intensity (FWHM) of
      $0.184 \pm 0.009$\,\AA.  The inferred line equivalent width is
      $W = 0.045 \pm 0.009$\,\AA. If we associate this feature with
      Na~I $\lambda$5890 (the stronger of the two components in the
      doublet), the observed wavelength is offset from the rest
      wavelength by $\Delta v = 194$\,km\,s$^{-1}$.  Similarly, the
      line is blueshifted from the systemic velocity of M101 ($v = 241
      \pm 2$\,km\,s$^{-1}$;\cite{1991trcb.book.....D}) by $\Delta
      v_{M101} = -47$\,km\,s$^{-1}$.  Given the high Galactic latitude
      ($b = 59.8^{\circ}$), we consider it likely that the absorbing
      material originates in M101 and the total extinction to the SN
      is negligible.}}
\end{center}
\end{figure}

\clearpage
\thispagestyle{empty}

\begin{figure}[t]
\begin{center}
\includegraphics[width=6.5in]{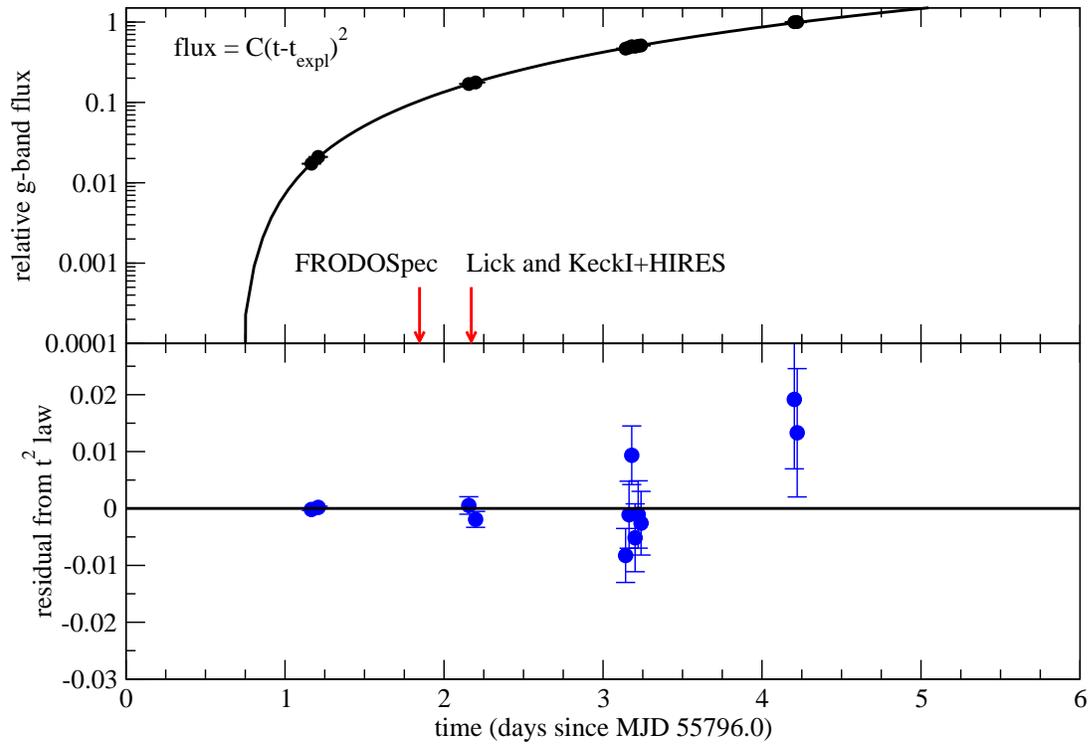}
  \caption{\label{fig:rise} \normalsize{Early photometry of SN~2011fe
      shows a parabolic rise and constrains the time of explosion.
      {\it Top} The relative $g$-band flux as a function of time for
      the first 4 nights after detection. Here we have fit the rise
      with a $t^2$ fireball model. {\it Bottom} The residuals from the
      fit.  Letting the exponent vary from 2, allowing for a potential
      departure from the fireball model, and only fitting the first 3
      nights of data, we find a best-fit explosion date of August
      23.687 $\pm$ 0.014 UT.  Based on these fits, our first data
      points were obtained just over 11\,hr after SN~2011fe exploded.
      All error bars presented are one standard deviation.} }
\end{center}
\end{figure}

\clearpage
\thispagestyle{empty}

\begin{figure}[t]
\begin{center}
\includegraphics[width=6.5in]{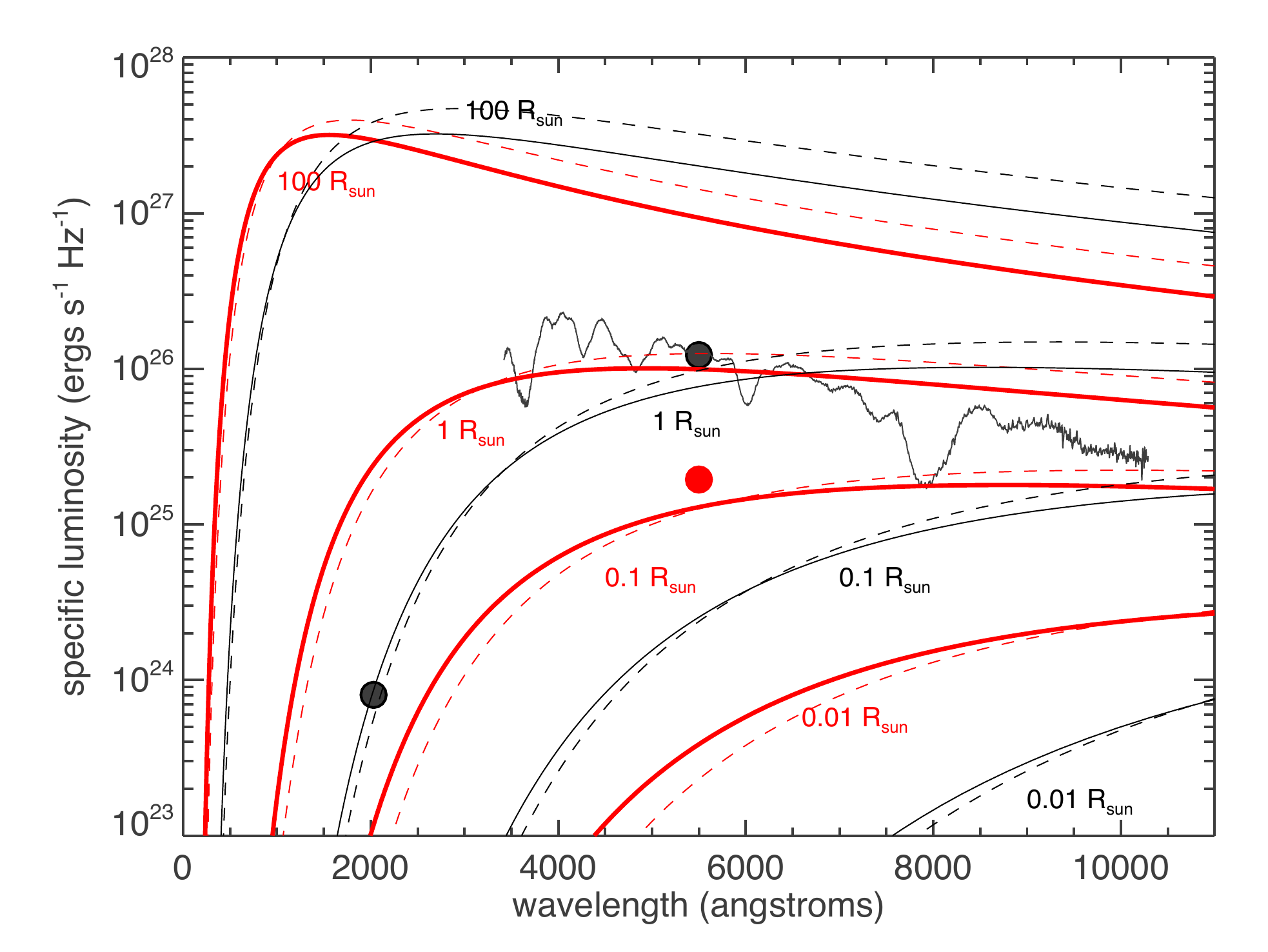}
  \caption{\label{fig:early_lc} \normalsize{Models and early data
      limit the radius of the exploding star of SN~2011fe.  The red
      lines show blackbody models for the day 0.5 spectrum assuming
      different values of the progenitor star radius $R_0$, where the
      solid lines and the dashed lines are derived from two
      separate analyses\cite{2010ApJ...708.1025K,2011arXiv1108.5548R}.
      The observed $g$-band photometry point at this time (red circle)
      constrains the radius of the progenitor star (or the surrounding
      opaque circumstellar medium) to be $\leq 10^{10}$\,cm
      $\approx$~0.1\,${\rm R}_\odot$.  The black lines and circles
      show corresponding model predictions and observations at day
      1.45.  In all models, we have assumed an ejecta mass equal to
      the Chandrasekhar mass, an explosion energy of $10^{51}$\,erg,
      and an opacity of 0.2\,cm$^2$ g$^{-1}$, appropriate for electron
      scattering in a singly ionised $A/Z = 2$ medium.  The early-time
      data indicate that the progenitor of SN~2011fe was a compact
      star, namely a white dwarf.  } }
\end{center}
\end{figure}

\clearpage
\pagenumbering{arabic}
\setcounter{page}{1}

\newcommand\figurenum[1]{%
 \def\thefigure{#1}%
 \addtocounter{figure}{1}%
}%

\def\simlt{\mathrel{\hbox{\rlap{\hbox{\lower4pt\hbox{$\sim$}}}\hbox{$<$}}}}
\def\simgt{\mathrel{\hbox{\rlap{\hbox{\lower4pt\hbox{$\sim$}}}\hbox{$>$}}}}

\def\ale{\mathrel{\hbox{\rlap{\hbox{\lower4pt\hbox{$\sim$}}}\hbox{$<$}}}}
\def\age{\mathrel{\hbox{\rlap{\hbox{\lower4pt\hbox{$\sim$}}}\hbox{$>$}}}}
\def\farcs{\hbox{$.\!\!^{\prime\prime}$}}
\def\farcm{\hbox{$.\mkern-4mu^\prime$}}
\def\fs{\hbox{$.\!\!^{\rm s}$}}

\def\nodata{---}

\def\ra#1#2#3{#1$^{\rm h}$#2$^{\rm m}$#3$^{\rm s}$}
\def\dec#1#2#3{$#1^\circ#2'#3''$}

\def\spose#1{\hbox to 0pt{#1\hss}}
\newcommand\lsim{\mathrel{\spose{\lower 3pt\hbox{$\mathchar"218$}}
     \raise 2.0pt\hbox{$\mathchar"13C$}}}
\newcommand\gsim{\mathrel{\spose{\lower 3pt\hbox{$\mathchar"218$}}
     \raise 2.0pt\hbox{$\mathchar"13E$}}}
\newcommand\ion[2]{#1$\;${\small\rmfamily\@Roman{#2}}\relax}
\newcommand{\vf}{\ensuremath{v_{\rm f}}}
\newcommand{\Nifs}{\ensuremath{^{56}\mathrm{Ni}}}
\newcommand{\Esn}{\ensuremath{E_{\rm sn}}}
\newcommand{\Eint}{\ensuremath{E_{\rm int}}}
\newcommand{\eint}{\ensuremath{\epsilon_{\rm int}}}
\newcommand{\Eth}{\ensuremath{E_{\rm th}}}
\newcommand{\Mej}{\ensuremath{M_{\rm ej}}}
\newcommand{\Msn}{\ensuremath{M}}
\newcommand{\Mni}{\ensuremath{M_{\rm ni}}}
\newcommand{\Msun}{\ensuremath{{\rm M}_{\odot}}}
\newcommand{\Lp}{\ensuremath{L_{\rm Ni}}}
\newcommand{\Ep}{\ensuremath{E_{\rm Ni}}}
\newcommand{\tp}{\ensuremath{\tau_{\rm Ni}}}
\newcommand{\td}{\ensuremath{t_{\rm d}}}
\newcommand{\te}{\ensuremath{t_{\rm e}}}
\newcommand{\Le}{\ensuremath{L_{\rm e}}}

\section{Supplementary Information}

\subsection{Low-Resolution Optical Spectroscopy:}

Our first spectra of SN~2011fe were obtained with the FRODOSpec
instrument on the Liverpool Telescope (La Palma, Canary Islands,
Spain). This dual-beam integral-field unit (IFU) spectrograph gives a
resolving power of 2200.  The supernova (SN) was observed on August
24, starting at 20.7 (UT dates are used throughout) with an exposure
time of 1800\,s.  The wavelength coverage is 3900--5600\,\AA\ in the
blue arm and 5900-9000\,\AA\ in the red arm.  The raw spectra for each
fiber of the IFU were extracted using a custom-built pipeline.  This
resulted in a sky-subtracted and wavelength-calibrated spectrum for
each fiber. They were then combined together and a flux calibration
was applied.

We also obtained low-resolution optical spectra of SN~2011fe with the
Kast spectrograph\cite{ms93} mounted on the 3\,m Shane telescope at
Lick Observatory beginning at 04:05 on 2011 August 25. It was taken
with a $2''$ wide slit, and a 600/3410 grism on the blue side and a
300/7500 grating on the red side, resulting in full width at
half-maximum (FWHM) resolutions of $\sim 4$ and 6\,\AA, respectively.
The spectrum was aligned along the parallactic angle to reduce
differential light losses\cite{f82}.  All spectra were reduced using
standard techniques.  Routine CCD processing and spectrum extraction
were completed with IRAF, and the data were extracted with an optimal
algorithm\cite{h86}. We obtained the wavelength scale from low-order
polynomial fits to calibration-lamp spectra.  Small wavelength shifts
were then applied to the data after cross-correlating a template sky
to the night-sky lines that were extracted with the SN.  We fit
spectrophotometric standard-star spectra to the data in order to flux
calibrate our spectra and to remove telluric lines\cite{wh88,mfh+00}.

The results of these observations are plotted in Figure~2 of the main
text.

\subsection{High-Resolution Optical Spectroscopy:}

We observed SN~2011fe with the High Resolution Echelle Spectrometer
(HIRES\cite{vab+94}) mounted on the 10\,m Keck I telescope beginning
at 5:51 on 2011 August 25 (only 1.6\,day after our derived explosion
date).  The spectrum was reduced using standard techniques, and
normalised by fitting the continuum in each echelle order with
low-order polynomials.  Wavelength calibration was performed relative
to a series of emission-line lamps, and then corrected to the Solar
System barycenter frame of reference.  A cutout of the resulting
normalised spectrum, centred on the rest-frame Na~I
$\lambda\lambda$5890, 5896 doublet, is shown in Figure~2 of the main
text.

In this wavelength range, we identify only a single significant
absorption feature.  Fitting a Gaussian profile to this line, we
measure a central wavelength of $\lambda = 5893.75 \pm 0.02$\,\AA\ and
a FWHM of $0.184 \pm 0.009$\,\AA.  The inferred line equivalent width
is $W = 0.045 \pm 0.009$\,\AA.

If we associate this feature with Na~I $\lambda$5890 (the stronger of
the two components in the doublet), the observed wavelength is offset
from the rest wavelength by $\Delta v = +194$\,km\,s$^{-1}$.
Similarly, the line is blueshifted from the systemic velocity of M101
($v = 241 \pm 2$\,km\,s$^{-1}$)\cite{ddc+91} by $\Delta v =
-47$\,km\,s$^{-1}$.  Given the high Galactic latitude ($b =
59.8^{\circ}$), we consider it likely that the absorbing material
originates in M101.

Independent of the origin of the absorbing material, the lack of
strong Na~I features indicates that the line of sight to SN~2011fe is
affected by a negligible amount of extinction.  Using common scaling
relations\cite{mz97}, the observed equivalent width corresponds to an
optical extinction of $A_{V} = 0.04$\,mag.

\subsection{\textit{Swift} UltraViolet/Optical Observations:}

Immediately upon discovery we triggered target-of-opportunity
observations of SN~2011fe with the \textit{Swift}
satellite\cite{gcg+04}.  Observations with the UltraViolet-Optical
Telescope (UVOT\cite{rkm+05}) began at 22:08 on 24 August 2011.  We
retrieved the level 2 UVOT data for SN~2011fe from the \textit{Swift}
data archive. To increase the signal-to-noise ratio, we stacked the
images for each individual filter on a daily basis.  To remove
host-galaxy contamination underlying the SN location, we subtracted
pre-outburst images of M101 obtained by the UVOT.  We caution that,
due to the nonlinearity of the coincidence-loss correction, such a
technique can lead to modest systematic uncertainties in the SN flux,
particularly at high count rates\cite{bhi+09}.  The photometric
calibration was performed following standard recipes\cite{pbp+08}, and
the reported magnitudes are on the AB system\cite{og83}. The resulting
photometry is presented in Supplementary Table~1.

\begin{table}
  \begin{center}
  \begin{tabular}{lcccc}
  \hline
  MJD & Telescope/Instrument & Filter & Exp. Time (s) & Magnitude \\
  \hline
55795.199 & P48 & $g$ & 120.0 & $> 21.5$ \\
55797.166 & P48 & $g$ & 60.0 & $17.349 \pm 0.011$ \\
55797.209 & P48 & $g$ & 60.0 & $17.155 \pm 0.011$ \\
55797.923 & \textit{Swift}/UVOT & $uvw1$ & 314.49 & $18.90 \pm 0.21$ \\
55797.925 & \textit{Swift}/UVOT & $u$ & 157.04 & $16.68 \pm 0.05$ \\
55797.926 & \textit{Swift}/UVOT & $b$ & 157.02 & $15.42 \pm 0.09$ \\
55797.927 & \textit{Swift}/UVOT & $uvw2$ & 629.48 & $20.96 \pm 0.39$ \\
55797.930 & \textit{Swift}/UVOT & $v$ & 157.02 & $15.12 \pm 0.09$ \\
55797.931 & \textit{Swift}/UVOT & $uvm2$ & 3264.03 & $> 21.0$ \\
55798.156 & P48 & $g$ & 60.0 & $14.886 \pm 0.010$ \\
55798.199 & P48 & $g$ & 60.0 & $14.839 \pm 0.009$ \\
55799.001 & \textit{Swift}/UVOT & $uvw1$ & 618.60 & $17.35 \pm 0.08$ \\
55799.002 & \textit{Swift}/UVOT & $u$ & 206.69 & $15.11 \pm 0.03$ \\
55799.003 & \textit{Swift}/UVOT & $b$ & 206.63 & $13.86 \pm 0.06$ \\
55799.003 & \textit{Swift}/UVOT & $uvw2$ & 1037.48 & $19.02 \pm 0.26$ \\
55799.006 & \textit{Swift}/UVOT & $v$ & 276.45 & $13.62 \pm 0.06$ \\
55799.006 & \textit{Swift}/UVOT & $uvm2$ & 1387.15 & $20.04 \pm 0.29$ \\
55799.142 & P48 & $g$ & 30.0 & $13.787 \pm 0.011$ \\
55799.164 & P48 & $g$ & 30.0 & $13.751 \pm 0.013$ \\
55799.181 & P48 & $g$ & 30.0 & $13.713 \pm 0.011$ \\
55799.202 & P48 & $g$ & 30.0 & $13.726 \pm 0.013$ \\
55799.221 & P48 & $g$ & 30.0 & $13.701 \pm 0.013$ \\
55799.239 & P48 & $g$ & 30.0 & $13.689 \pm 0.012$ \\
55800.203 & P48 & $g$ & 30.0 & $12.964 \pm 0.013$ \\
55800.221 & P48 & $g$ & 30.0 & $12.959 \pm 0.012$ \\
  \hline
  \end{tabular}
  \end{center}
  \caption{UV/Optical Observations of SN~2011fe.    
    P48 observations have been calibrated with respect to Sloan Digital
    Sky Survey $g$-band images of the field, and are on the PTF photometric
    system.  \textit{Swift}/UVOT images have been calibrated using
    standard recipes\cite{pbp+08} and are reported on the AB system\cite{og83}.}
\end{table}

\subsection{SYNAPPS Spectral Fits:}

We analysed the spectrum of SN~2011fe using the automated SN spectrum
interpretation code SYNAPPS\cite{2011PASP..123..237T}.  SYNAPPS uses a
parallelised pattern search algorithm to compare highly parametrised
synthetic spectra to observed ones to find a good fit.  Results from
SYNAPPS are useful for (a) identifying or rejecting the presence of
ion signatures and (b) estimating characteristic ejecta velocities.
Typical premaximum SN~Ia ions are confirmed: O~I, Mg~II, Si~II, S~II,
Ca~II, and Fe~II (Supplementary Figure 1).  Fe~III was not needed in
the fit.  In addition, the fit confirms the presence of C~II
$\lambda\lambda$6580, 7234, though the detailed fit is not perfect,
probably due to parametrisation bias.  Both high-velocity (HV) Si~II
and Ca~II are confirmed by SYNAPPS (extending above 21,000 km/s).
Surprisingly, SYNAPPS finds HV O~I (in excess of 20,000 km/s) for the
absorption centered at 7400~\AA.  This is the first identification of
variable high-velocity oxygen in the ejecta of a SN~Ia in the
literature to date. Furthermore we note that the change in velocity of
the Mg~II and Fe~II features are $\sim$4\% and that of Si~II is
$\sim$8\% over the first 24\,hr.

\begin{figure}
\centering
\includegraphics[width=6.4in,angle=0]{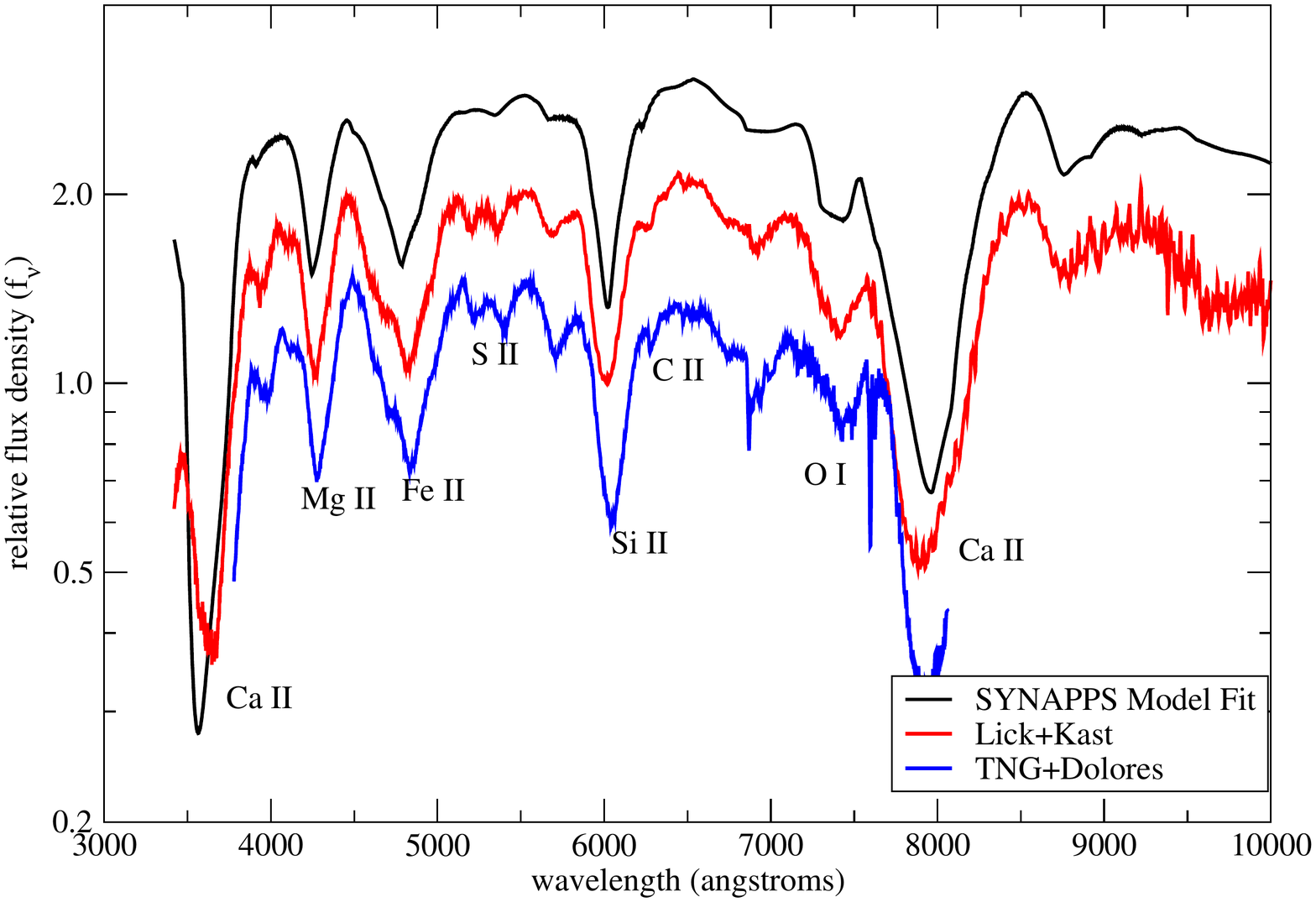}
\caption{A SYNAPPS fit to the Lick/Kast spectrum and the TNG/Dolores
  spectrum taken 16 hrs later. The agreement is excellent and we have
  confidently identified several intermediate-mass elements including
  high-velocity Ca~II, Si~II, and O~I, the last of which forms the
  feature centered at 7400~\AA.}
\end{figure}

\subsection{Parabolic Shape of the  \Nifs\ Powered Light Curve at Early Times:}

The $L \propto t^2$ behaviour found for the early-time light curve of
SN~2011fe is consistent with a simple fireball model in which the
effective temperature remains fixed while the radius increases with
time as $r = v_p t$.  In reality, one expects the effective
temperature to change with time, while the photospheric velocity $v_p$
will decrease as the remnant expands and the density drops.  However,
the same $t^2$ law can be derived by a more rigorous analytic argument
that includes the effects of radiative diffusion, \Nifs\ energy
deposition, and expansion losses\cite{a82}.  We describe here a simple
one-spatial-zone model for the whole remnant which shows that the
observed rise, $L \propto t^2$ at early times, is consistent with a
\Nifs\ powered SN, with no contribution to the luminosity from the
explosion shock wave or from interaction with circumstellar material
or a companion star.

For the expanding remnant, the evolution of the internal energy,
\Eint, is given by the first law of thermodynamics,
\begin{equation}
\frac{ \partial \Eint}{\partial t} =
- P \frac{ \partial  V}{\partial t}  + \Lp(t) -  \Le(t),
\label{Eq:first_law}
\end{equation}
where \Lp\ is the energy deposited per second from \Nifs\ decay and
\Le\ is the radiated luminosity.  We assume that the \Nifs\ energy is
thermalised throughout the remnant, and that radiation pressure
dominates, $P = \Eint/3V$. Assuming homologous expansion, the volume
increases as $V \propto t^3$ and equation~\ref{Eq:first_law} becomes
\begin{equation}
\frac{1}{t} \frac{ \partial } {\partial t} [\Eint t]= \Lp(t) - \Le(t).
\label{Eq:diff_Eint}
\end{equation}
The radiated luminosity, \Le, is approximated from the diffusion
equation
\begin{equation}
{\Le \over 4 \pi R^2} =   \frac{c}{3 \kappa \rho} \frac{\partial \Eint/V}{\partial r}
\approx \frac{c}{ 3\kappa \rho}  \frac{\Eint/V}{R}.
\end{equation}
In homologous expansion, $R = \vf t$, and the radiated luminosity can
be rewritten
\begin{equation}
\Le = \frac{\Eint t}{\td^2},~~{\rm where}~~\td = \biggl[ \frac{3}{4 \pi} \frac{ \Mej \kappa }{\vf c}  \biggr]^{1/2}.
\label{Eq:Le}
\end{equation}
Here, we defined the effective diffusion time \td, and $\vf =
[\Esn/2\Mej]^{1/2}$ is the final characteristic ejecta velocity.  For
the case where the elapsed time is much less than the \Nifs\ decay
time of $\tp = 8.8$~day, the energy deposition from \Nifs\ decay can
be considered constant with time ($\Lp = \Ep/\tp$) and the solution of
equation~\ref{Eq:Le} is
\begin{equation}
\Le(t) \approx \frac{\Ep}{\tp} [1 - e^{-t^2/2 \td^2} ]~~~~~~~~~~~~ t \ll \tp ,
\end{equation}
where we have assumed that the initial internal energy of the ejecta
(from the explosion shock) is negligible compared to the heating from
\Nifs\ decay.  For typical parameters, the effective diffusion time is
$\sim 20$~day, so taking the limit $t \ll \td$ gives
 \begin{equation}
\Le(t) \approx \frac{\Ep}{\tp} \frac{t^2}{2 \td^2}~~~~~~~~~~~~ t \ll \td, t \ll \tp .
\label{eq:Le}
 \end{equation}

 The validity of the argument can be questioned at these very early
 times, as the analysis does not properly capture the early transient
 phase in which a diffusion wave moves from the inner \Nifs\ core to
 the surface.  However, the excellent fit of the observed light curve
 to a $t^2$ law suggests that the simple model is not unreasonable.
 From the day~0.45 observation of SN~2011fe (luminosity $L \approx
 10^{40}\,{\rm erg~s^{-1}}$) and equation~\ref{eq:Le}, we find a total
 \Nifs\ mass of $0.45~\Msun$, comparable to the value inferred for a
 (slightly underluminous) normal SN~Ia.  This suggests that the
 early-time luminosity of SN~2011fe is consistent with \Nifs\ powering
 only, with little or no contribution from shocks.

\subsection{Contribution to the Early Luminosity from Shock Heating:}

At early times ($\sim 1$~day), the diffusion of radiation from the
ejecta shock-heated in the explosion may contribute to the emergent
luminosity (in addition to the luminosity generated by \Nifs\
heating).  Observations of this shock luminosity can be used to
constrain the radius of the progenitor star.  Relevant models have
been considered by many
authors\cite{chevalier_1992,chevalier_2008,piro_2010,kasen_2010,nakar_2011,rabinak_2011a,rabinak_2011b},
all of whom take a similar analytical approach to calculating the
evolution of a radiation dominated, homologously expanding, constant
opacity, spherically symmetric supernova remnant.  The models differ
in their assumptions of the initial ejecta density and pressure
profiles and in the treatment of radiative diffusion.  However, the
final predictions of the early-time light curve tend to be quite
similar.

Recent work has shown\cite{rabinak_2011b}, for the case of early
cooling luminosity after shock breakout in a SN~Ia, the luminosity and
effective temperature follow:
\begin{equation}
\begin{split}
L(t) &= 1.2 \times  R_{10} 10^{40} E_{51}^{0.85} M_c^{-0.69} \kappa_{0.2}^{-0.85} f_p^{-0.16} t_d^{-0.31} ~{\rm erg~s^{-1}},\\
T(t) &=
4015 ~ R_{10}^{1/4} E_{51}^{0.016}  M_c^{0.03} \kappa_{0.2}^{0.27} f_p^{-0.022}  t_d^{-0.47}~{\rm K},
\end{split}
\label{eq:rabinak}
\end{equation}
where $E_{51}$ is the explosion energy $E/10^{51}$~erg, $R_{10}$ is
the progenitor radius $R/10^{10}$~cm, $M_c$ is the total ejecta mass
in units of the Chandrasekhar mass, $\kappa_{0.2}$ is the opacity
$\kappa/0.2~{\rm cm^{2}~g^{-1}}$, $f_p$ is a dimensionless form
factor, and $t_d$ is the time since explosion in days.  These results
are similar to those found by earlier work\cite{piro_2010} for ejecta
in spherical expansion.

For the case of SN ejecta impacting a companion star in Roche-lobe
overflow, self-similar diffusion arguments have shown that the post
interaction luminosity is described by\cite{kasen_2010}:
\begin{equation}
\begin{split}
L(t) &= 1.0 \times 10^{40} a_{10}~ E_{51}^{0.875}  M_c^{-0.375} \kappa_{0.2}^{-0.75} t_d^{-0.5} ~{\rm erg~s^{-1}},\\
T(t) &=
4446 ~ a_{10}^{1/4} E_{51}^{0.0}  M_c^{0.0} \kappa_{0.2}^{0.97}  t_d^{-0.51}~{\rm K},\end{split}
\label{eq:kasen}
\end{equation}
where $a_{10}$ is the separation distance between the WD and its
companion star in units of $10^{10}$\,cm.  The isotropic equivalent
luminosity here applies to a viewing angle aligned with the symmetry
axis, and will be lower for off-axis viewing angles. Despite the
somewhat different context and approach, the formulae are very similar
to the previously discussed colling of the shock heated
SN\cite{rabinak_2011b}, both in the scalings and the overall
normalization.  In particular, the luminosity depends linearly on $R$
or $a$, while the effective temperature is proportional to $R^{1/4}$
or $a^{1/4}$.

We used equations~\ref{eq:rabinak} and \ref{eq:kasen} to construct
model spectra for progenitors of different radii.  We assumed values
typical of SN~Ia models, namely $E_{51} = 1, M_c = 1, \kappa_{0.2} =
1, f_p = 1$.  The emission spectrum was taken to be a blackbody with
temperature $T$ and luminosity $L$.  The day 0.45 $g$-band observation
of SN~2011fe indicates $L \sim 10^{40}~{\rm erg~s^{-1}}$, which
constrains the progenitor radius to $R \leq 10^{10}$~cm.

Because the impact with a companion star only shocks a portion of the
SN ejecta (that within a conical region with opening angle $\sim
40^\circ$), the shock luminosity from the interaction is anisotropic,
and will be most prominent for viewing angles nearly aligned with the
symmetry axis.  Such an orientation occurs $\sim 10\%$ of the time.
Numerical multi-dimensional radiation transport calculations of the
dependence of the luminosity on viewing angle\cite{kasen_2010} show
that the luminosity observed $90^\circ$ ($180^\circ$) from the
symmetry axis is about a factor $10~(100)$ lower than that viewed
on-axis.  For a red-giant companion ($a \approx 10^{13}$~cm), the
predicted shock luminosity is $\geq 10^{41}\,{\rm erg~s^{-1}}$ for all
viewing angles, ruling out this progenitor system.  A 1\,\Msun\
main-sequence companion ($a \approx 10^{11}$\,cm) has $L \approx
10^{41}\,{\rm erg~s^{-1}}$ when viewed on-axis, and thus would be
consistent with the data if the observer were oriented $\geq 90^\circ$
from the symmetry axis.

\end{document}